\documentclass[10pt,a4paper]{article}
\pdfoutput=1

\usepackage{jheppub} \usepackage{bm,float} \usepackage[caption = false]{subfig}
\usepackage[section]{placeins} 
\usepackage{array, multirow, booktabs, slashed, amsthm, xfrac, fix-cm}
\usepackage{url, hyperref, amsmath, amssymb}
\usepackage{parskip, setspace}

\usepackage{etoolbox}
\makeatletter
\patchcmd{\maketitle}{\@fpheader}{ }{ }{ }
\makeatother

\def \Qbar{\overline{Q}}  \def \Lbar{\overline{L}} \def
\ubar{\overline{u}} \def \dbar{\overline{d}} \def \ebar{\overline{e}} 
\def \Htilde{\widetilde{H}} 
\def \LC{\epsilon}

\newtheorem{mydef}{Rule}

\title{Extending the Standard Model Effective Field Theory with the Complete Set
	of Dimension-7 Operators} 
\author{Landon Lehman}
\affiliation{Department of Physics \\
University of Notre Dame \\ Notre Dame, IN 46556} \emailAdd{llehman@nd.edu}
\date{\today}
\preprint{ }

\abstract{We present a complete list of the independent dimension-7 operators
	that are constructed using the Standard Model degrees of freedom and are
	invariant under the Standard Model gauge group. 
	This list contains only 20 independent operators; far fewer than the 63
	operators available at dimension 6.
	All of these dimension-7 operators contain fermions and violate lepton
	number, and 7 of the 20
	violate baryon number as well.
	This result extends the Standard Model Effective Field Theory
	and allows a more detailed exploration of the structure and
	properties of possible deformations from the Standard Model
	Lagrangian.}

\begin{document}

\maketitle

\section{Introduction} 

The discovery of the Higgs boson at the Large Hadron Collider (LHC)
provided the latest of many examples of the explanatory power of the Standard
Model. However, because of the hierarchy problem and the accompanying angst
regarding naturalness, it is widely believed that
beyond-the-Standard-Model (BSM) physics is necessary at scales not much beyond
the TeV scale. As of yet there are no clear experimental
signatures of new physics at these scales; in particular, theoretically compelling
extensions of the Standard Model such as supersymmetry have yet to be
experimentally verified. 

While searches for BSM particles at the LHC driven by explicit theories such as
supersymmetry or extra dimensions are useful and ongoing, it may be beneficial
to take an alternative and complementary approach in the quest for BSM physics.
This avenue is the method of effective field theory, which parametrizes all
possible deviations from the Standard Model that any particular UV theory might
explore. Effective field theory can be viewed as a universal bottom-up approach to BSM
physics, as opposed to the top-down approach of particular UV theories.

Assuming no
undiscovered light ($\lesssim$ TeV) particles, integrating out the heavy degrees
of freedom in a BSM model will produce effective operators that are invariant
under the Standard Model gauge group. By constructing every possible operator
using the Standard Model degrees of freedom and using Standard Model gauge group
invariance as a constraint, we can remain agnostic as to the specific BSM theory
that is producing these operators. Of course, a given UV theory may have
symmetries that forbid some of the operators, or various operators may be
loop suppressed, but the point of effective field theory is to take a
general approach and cast a wide net. The operators constructed in this manner collectively
constitute the Standard Model Effective Field Theory (SMEFT).

\begin{table}
\begin{center} { 
\begin{tabular}{cc} 
	\toprule Dimension & Number of operators \\ 
	\midrule 	2 & 1 \\ 
			4 & 13 \\ 
			5 & 1 \\ 
			6 & 63 \\ 
			7 & 20 \\ 
	\bottomrule 
\end{tabular}}
\caption{The number of operators invariant under the Standard Model
	gauge group $SU(3)_C\otimes SU(2)_W\otimes U(1)_Y$, organized by
	canonical operator dimension. 
	This enumeration only includes
	operators with nontopological effects and does not consider flavor
	permutations or Hermitian conjugates.
	Note
	that four of the dimension-6 operators violate baryon number
	conservation, leaving 59 that conserve baryon number. } 
\end{center} 
\label{numbers}
\end{table}

Since operators with dimension greater than 4 are suppressed by
powers of an
energy scale equal to the scale at which the new physics is integrated out, the
construction of effective operators from Standard Model fields
can be organized by an expansion in the canonical dimension of the
operators. The canonical dimension is the total dimension of the fields making
up an operator; dimensionful couplings are not
included. This expansion can be treated as an expansion in powers of the dimensionless
parameter
$\varepsilon = m_{\text{SM}}/\Lambda$, 
where $\Lambda$ is the scale of the new physics and 
Standard Model mass scales such as the Higgs mass or the top quark mass are
represented by $m_\text{SM}$.

This constructive program is trivially
implemented at dimension 5, or equivalently $\varepsilon^1$, since there is only one possible
dimension-5 gauge invariant operator - the Weinberg neutrino mass operator
\cite{Weinberg:1979sa}. At the dimension-6 level ($\varepsilon^2$),
Buchm{\"u}ller and Wyler
counted 80 operators in 1986 \cite{Buchmüller1986621}. Some of these 80
operators were redundant and able be interrelated by using the Standard Model equations
of motion to make field redefinitions. An updated classification containing 59
independent dimension-6 operators was published in 2010
\cite{Grzadkowski:2010es}.  Some dimension-7 and dimension-8 operators have been
studied in the literature \cite{Degrande:2013kka, Weinberg:1980bf,
Weldon:1980gi, Babu:2012iv, Babu:2012vb, Chalons:2013mya, delAguila:2012nu,
Babu:2001ex, Bonnet:2009ej, deGouvea:2007xp, Angel:2012ug}, but no complete
operator basis has previously been published for any dimension greater than 6.

Table \ref{numbers} lists the number of operators in the SMEFT
for each order in the operator dimension expansion up to dimension 7. This
operator count does not
include flavor index
permutations or Hermitian conjugates and ignores operators that only contribute
to topological
quantum effects, such as the QCD theta term $\theta
g_3^2 G^A_{\mu\nu}\widetilde{G}^{A\mu\nu}/(32\pi^2)$.
The Higgs mass term $(H^\dagger H)$ is the solitary dimension-2 operator.

The formal properties of the dimension-6 operators of the SMEFT have been extensively
studied. A series
of papers calculated the full $59\times 59$ anomalous dimension matrix for the
operators that preserve baryon number
\cite{Grojean:2013kd, Jenkins:2013zja, Jenkins:2013wua, Alonso:2013hga,
	Elias-Miro:2013gya, Elias-Miro:2013mua, Elias-Miro:2013eta}, and the
anomalous dimension matrix for the baryon-number violating operators followed
soon after \cite{Alonso:2014zka}.\footnote{This reference also noted that only
	four
of the five baryon-number violating operators listed in Ref.~\cite{Grzadkowski:2010es}
are independent, a fact previously realized in Ref.~\cite{Abbott:1980zj}.}
The dimension-6 operators were also recently noted to have intriguing
``holomorphic'' properties \cite{Alonso:2014rga}. Similar examinations of the
formal properties of the dimension-7
operators can now be undertaken utilizing the list presented in this paper.

The SMEFT up to dimension 6 has also been widely used in phenomenological
studies. The LHC phenomenology of a specific toy model leading to some
dimension-6 operators was examined in Ref.~\cite{deVries:2014apa}.
The effects of the dimension-6 SMEFT on lepton flavor violation
\cite{Crivellin:2013hpa} and top quark processes \cite{Zhang:2010dr,
Bach:2014zca} have also been
explored. Many
papers address the effects of dimension-6 operators on
the Higgs sector; for a representative sample see Refs.~\cite{Bonnet:2011yx,
	Brivio:2013pma, Dumont:2013wma, 
Contino:2013kra, Masso:2014xra, Bonnet:2012nm, Boos:2013mqa, Gavela:2014vra} and
references therein.
The implementation of operators involving the Higgs into FeynRules has aided
such
phenomenological efforts, allowing dimension-6 operators to be used in Monte
Carlo generators such as MadGraph \cite{Alloul:2013naa}.
The dimension-7 operators will also be useful for phenomenological studies of
possible signals of new physics in various channels and processes. In
particular, given the lepton and baryon-number violating properties of the
dimension-7 operators, they can be used to explore baryogenesis and
leptogenesis; giving possible methods of generating the
matter-antimatter asymmetry of the Universe.

The organization of this paper is as follows. Section \ref{notation} establishes
the relevant symbols and conventions, and
the complete list of the independent
dimension-7 operators is presented in Section \ref{list}. Some necessary facts
regarding fermions and hypercharge are reviewed in Section \ref{fermions_hyper}.
In Section \ref{classification} we work through the details of the operator classification
used to obtain the list in Section \ref{list}, and we conclude in Section
\ref{conclusion}.

\section{Notation and conventions}
\label{notation}

\begin{table}
	\begin{center}
		{
		\begin{tabular}{cccccc}
		\toprule
		Field	& $SU(3)_C$ & $SU(2)_W$ & $U(1)_Y$ & dimension &
		$SL(2,\mathbb{C})$\\
		\midrule
		$Q_L = \begin{pmatrix} u_L \\ d_L \end{pmatrix}$ & $\bm{3}$ &
		$\bm{2}$ & 1/6 & 3/2 & $\left( \frac{1}{2},
		0 \right)$ \\
		\midrule
		$u_R$ 	& $\bm{3}$ & $\bm{1}$ & 2/3 & 3/2 &
		$\left( 0, \frac{1}{2} \right)$ \\
		\midrule
		$d_R$	& $\bm{3}$ & $\bm{1}$ & -1/3 & 3/2 &
		$\left(0, \frac{1}{2} \right)$ \\
		\midrule
		$H = \begin{pmatrix} H^+ \\ H^0 \end{pmatrix}$  & $\bm{1}$ &
		$\bm{2}$ & 1/2 & 1 & $\left( 0, 0 \right)$ \\
		\midrule
		$L_L = \begin{pmatrix}\nu_L \\ e_L \end{pmatrix}$ & $\bm{1}$ &
		$\bm{2}$ & -1/2 & 3/2 & $\left( \frac{1}{2},
		0 \right)$ \\
		\midrule
		$e_R$ & $\bm{1}$ & $\bm{1}$ & -1 & 3/2 & $\left( 0,
		\frac{1}{2} \right)$ \\
		\bottomrule
	\end{tabular}}
	\caption{The Standard Model matter degrees of freedom, along with their
	dimensions and gauge and Lorentz group representations.}
	\label{matter_fields}
	
\end{center}
\end{table}

\begin{table}
	\begin{center}
		{
			\begin{tabular}{cccc}
			\toprule
			Field	& $SU(3)_C$ & $SU(2)_W$ & $U(1)_Y$ \\
			\midrule
			$G^A_\mu$ & $\bm{8}$ & $\bm{1}$ & $0$  \\
			\midrule
			$W^I_\mu$ & $\bm{1}$ & $\bm{3}$ & $0$ \\
			\midrule
			$B_\mu$ & $\bm{1} $ & $\bm{1}$ & $0$  \\
		        \bottomrule
		\end{tabular}}
		\caption{The Standard Model gauge degrees of freedom and their
			gauge group representations. All have
			dimension 1 and transform in the vector representation
			of the Lorentz group. $A =
		1,\dots, 8$ and $I = 1,\dots ,3$.}
		\label{gauge_fields}
	\end{center}
\end{table}	

The Standard Model degrees of freedom are listed for convenience in Tables
\ref{matter_fields}
and \ref{gauge_fields}.
The $SU(2)_W$ generators will be
represented by $\tau^I$, with $I=1,2,3$ the adjoint representation indices. The
indices for the fundamental representation of
$SU(2)_W$ will be $\{i,j,k,m,n\} \in \{1,2\}$.
The field-strength tensors will be denoted by $X_{\mu\nu} \in \{G^A_{\mu\nu},
W^I_{\mu\nu}, B_{\mu\nu}\}$, all with dimension 2. Dual tensors are defined as
$\widetilde{X}_{\mu\nu} = (1/2)\LC_{\mu\nu\rho\sigma}X^{\rho\sigma}$. 
The indices $\{p, r, s, t\}$ will be used to denote fermion flavors
(generations), and the chirality indices $\{L,R\}$ will generally be
suppressed. To satisfy $SU(2)_W$ invariance, the complex conjugate of
the Higgs appears in the construction
$\Htilde^i \equiv \LC_{ij}(H^j)^*$. The symbol $C$ will denote the Dirac charge
conjugation matrix, which links together same-chirality fermion fields in a
scalar current.

Color $SU(3)_C$ indices will always be suppressed, with the convention that an
operator with two
quarks with color indices $\{\alpha,\beta\}$ will always be contracted as
$\delta_{\alpha\beta}q^\alpha q^\beta$, and an operator with three quarks will
have the color indices contracted in the totally antisymmetric manner
$\LC_{\alpha\beta\gamma}q^\alpha q^\beta
q^\gamma$. There are no dimension-7 operators with more than three quarks.

The SMEFT Lagrangian can contain the
matter fields shown in Table \ref{matter_fields},
the gauge field-strength tensors $X_{\mu\nu}$, and covariant derivatives
$D_\mu$.
The SMEFT Lagrangian at zeroth order, otherwise known as the Standard
Model Lagrangian, is
\begin{equation}
\begin{split}
	\mathcal{L}_{SM} & = -\frac{1}{4}G_{\mu\nu}^A G^{A\mu\nu} - \frac{1}{4} 
	W_{\mu\nu}^I W^{I\mu\nu} - \frac{1}{4} B_{\mu\nu}B^{\mu\nu} 
	 + \left(D_\mu H \right)^\dagger \left( D^\mu H \right) + m^2 H^\dagger H -
	\frac{1}{2} \lambda \left(H^\dagger H \right)^2 \\
	&+ i \left( \Lbar \slashed{D} L + \ebar \slashed{D}e +
	\Qbar \slashed{D}Q + \ubar \slashed{D}u +
	\dbar \slashed{D} d \right) 
	 - \left( \Lbar Y_e e H + \Qbar Y_u u \Htilde +
	\Qbar Y_d d H + \text{h.c.} \right) .
\end{split}
\label{sm}
\end{equation}
As mentioned before, we ignore topological terms - i.e. terms that are total
spacetime derivatives. The Lagrangian and all operators apply above the
electroweak symmetry breaking scale.

The dimension-7 operators presented in Section \ref{list} generically violate
lepton and/or baryon number. In fact, all SMEFT operators of odd dimension
must violate either lepton number or baryon number, and perhaps both
\cite{Degrande:2012wf, deGouvea:2014lva, Rao:1983sd}.
Baryon number is defined as
\begin{equation}
	B = \frac{1}{3}(n_q - n_{q^c}),
	\label{}
\end{equation}
where $n_q$ is the number of quarks and $n_{q^c}$ is the number of antiquarks.
Lepton number is
\begin{equation}
	L = (n_l - n_{l^c}),
	\label{}
\end{equation}
with $n_l$ the number of leptons and $n_{l^c}$ the number of antileptons.

\section{Complete list of dimension-7 operators}
\label{list}
\begin{table}[h]
\begin{center}
\begin{minipage}[t]{0.45\textwidth}
\renewcommand{\arraystretch}{1.5}
\begin{tabular}[t]{c|c}
\multicolumn{2}{c}{1 : $\psi^2 H^4$ + h.c.} \\
\hline
$ \mathcal{O}_{LH}$ 	& $\LC_{ij}\LC_{mn}(L^i CL^m)H^j H^n (H^\dagger H)$ \\
\end{tabular}
\end{minipage}
\begin{minipage}[t]{0.45\textwidth}
\renewcommand{\arraystretch}{1.5}
\begin{tabular}[t]{c|c}
\multicolumn{2}{c}{2 : $\psi^2 H^2 D^2$ + h.c.} \\
\hline
$ \mathcal{O}_{LHD}^{(1)} $ & $\LC_{ij}\LC_{mn}L^i C(D^\mu L^j)H^m (D_\mu H^n)$\\
$ \mathcal{O}_{LHD}^{(2)} $ & $\LC_{im}\LC_{jn}L^i C(D^\mu L^j)H^m (D_\mu H^n)$\\
\end{tabular}
\end{minipage}
\vspace{0.25cm}
\begin{minipage}[t]{0.45\textwidth}
\renewcommand{\arraystretch}{1.5}
\begin{tabular}[t]{c|c}
\multicolumn{2}{c}{3 : $\psi^2 H^3 D$ + h.c.} \\
\hline
$\mathcal{O}_{LHDe}$ 	& $\LC_{ij}\LC_{mn}\left(L^i C\gamma_\mu e\right) H^j H^m
D^\mu H^n$  \\
\end{tabular}
\end{minipage}
\begin{minipage}[t]{0.45\textwidth}
\renewcommand{\arraystretch}{1.5}
\begin{tabular}[t]{c|c}
\multicolumn{2}{c}{4 : $\psi^2 H^2 X$ + h.c.} \\
\hline
$\mathcal{O}_{LHB}$ 	& $\LC_{ij}\LC_{mn}\left(L^i C\sigma_{\mu\nu}L^m\right)
H^j H^n B^{\mu\nu}$ \\
$\mathcal{O}_{LHW}$ 	& $\LC_{ij}(\tau^I \LC)_{mn} \left(L^i C \sigma_{\mu\nu} L^m\right)
H^j H^n W^{I\mu\nu}$ \\
\end{tabular}
\end{minipage}
\vspace{0.25cm}
\begin{minipage}[t]{0.45\textwidth}
\renewcommand{\arraystretch}{1.5}
\begin{tabular}[t]{c|c}
\multicolumn{2}{c}{5 : $\psi^4 D$ + h.c.} \\
\hline
$\mathcal{O}_{LL\dbar uD}^{(1)}$ & $\LC_{ij}(\dbar \gamma_\mu u)(L^i C D^\mu
L^j)$ \\
$\mathcal{O}_{LL\dbar uD}^{(2)}$ & $\LC_{ij}(\dbar \gamma_\mu u)(L^i C
\sigma^{\mu\nu}D_\nu L^j)$ \\
$\mathcal{O}_{\Lbar QddD}^{(1)}$ & $(QC\gamma_\mu d)(\Lbar D^\mu d) $ \\
$\mathcal{O}_{\Lbar QddD}^{(2)}$ & $(\Lbar \gamma_\mu Q)(dCD^\mu d) $ \\
$\mathcal{O}_{ddd\ebar D} $	& $ (\ebar \gamma_\mu d)(dCD^\mu d)$ \\
\end{tabular}
\end{minipage}
\begin{minipage}[t]{0.45\textwidth}
\renewcommand{\arraystretch}{1.5}
\begin{tabular}[t]{c|c}
\multicolumn{2}{c}{6 : $\psi^4 H$ + h.c.} \\
\hline
$\mathcal{O}_{LLL\ebar H} 	$	& $\LC_{ij}\LC_{mn}(\ebar L^i)(L^j
CL^m)H^n$  \\
$\mathcal{O}_{LLQ\dbar H}^{(1)} $ & $\LC_{ij}\LC_{mn}(\dbar L^i)(Q^j CL^m)H^n$ \\
$\mathcal{O}_{LLQ\dbar H}^{(2)} $ & $\LC_{im}\LC_{jn}(\dbar L^i)(Q^j CL^m)H^n$
\\
$\mathcal{O}_{LL\Qbar uH} $	& $\LC_{ij}(\Qbar_m u)(L^m C L^i)H^j$	\\
$\mathcal{O}_{\Lbar QQdH}  	$	& $\LC_{ij}(\Lbar_m d)(Q^m C
Q^i)\Htilde^j$ \\
$\mathcal{O}_{\Lbar dddH} $ 	& $(dCd)(\Lbar d)H$ \\
$\mathcal{O}_{\Lbar uddH} $ 	& $(\Lbar d)(uCd)\Htilde$ \\
$\mathcal{O}_{Leu\dbar H} $ 	& $\LC_{ij}(L^i C\gamma_\mu e)(\dbar \gamma^\mu
u)H^j$ \\
$\mathcal{O}_{\ebar QddH} $ 	& $\LC_{ij}(\ebar Q^i)(dCd)\Htilde^j$ \\
\end{tabular}
\end{minipage}
\end{center}
\caption{The dimension-seven operators. Color and flavor indices are left
	implicit, and 
	$SU(2)_W$ indices are left implicit when the contractions are
	obvious. The symbol $C$ represents the Dirac charge conjugation matrix,
	as explained in Section \ref{fermions_hyper}. The six classes of
	operators shown group the operators by the degrees of freedom $H, X,
	D$, and $\psi$.}
\label{the_list}
\end{table}

The complete list of independent dimension-7 operators is presented in Table
\ref{the_list}.
It will be shown in Section \ref{fermions_hyper} that all of the possible dimension-7
operators contain fermions, so every operator in Table \ref{the_list} has
suppressed flavor indices. 
This can be contrasted with the dimension-6 case, where four of the eight
operator classes are fermion free and thus do not need flavor indices.
Some operators which might at first glance seem to be missing from the list can be
formed from those included by adding or subtracting combinations with different
permutations of flavor indices. For example, 
\begin{equation}
	\LC_{in}\LC_{jm} (\dbar L_p^i)(Q^j L_q^m) H^n = \mathcal{O}_{LLQ\dbar
	H}^{(2)pq} - \mathcal{O}_{LLQ\dbar H}^{(1)pq} .
	\label{}
\end{equation}
where the Schouten identity $\LC_{in}\LC_{jm} = \LC_{im}\LC_{jn} -
\LC_{ij}\LC_{mn}$ was used. Writing out the lepton doublet flavor
indices explicitly, we have 
\begin{equation}
	\mathcal{O}_{LLQ\dbar H}^{(1)pq} = \LC_{ij}\LC_{mn}(\dbar L_p^i)(Q^j C
	L_q^m) H^n \quad \text{and} \quad \mathcal{O}_{LLQ\dbar H}^{(2)pq} =
	\LC_{im}\LC_{jn} (\dbar L_p^i)(Q^j C L_q^m) H^n .
	\label{}
\end{equation}
Another more complicated example using the same two operators is 
\begin{equation}
	\LC_{in}\LC_{jm}(\dbar Q^j)(L_p^i L_q^m)H^n =
	-\left(\mathcal{O}_{LLQ\dbar H}^{(2)pq} +
	\mathcal{O}_{LLQ\dbar H}^{(2)qp}\right) + 
	\left(\mathcal{O}_{LLQ\dbar H}^{(1)pq} +
	\mathcal{O}_{LLQ\dbar H}^{(1)qp}\right) ,
	\label{}
\end{equation}
where Fierz identities were used along with the Schouten identity mentioned
above.
As a third example, Ref.~\cite{Weinberg:1980bf} lists two dimension-7 operators with
the field
content $\{\Lbar, Q,Q, d, H\}$, which after passing to the notation used here
are
\begin{equation}
	\mathcal{O}_1^{pq} = \LC_{ij}\LC_{mn}(Q_p^i C Q_q^j)(\Lbar^m
	d)\Htilde^n \quad \text{and}\quad \mathcal{O}_2^{pq} =
	\LC_{jm}\LC_{in}(Q^i_p C Q_q^j)(\Lbar^m d)\Htilde^n .
	\label{}
\end{equation}
In terms of the single operator $\mathcal{O}_{\Lbar QQdH}^{pq}$ given in
Table \ref{the_list}, these can be written
\begin{equation}
\begin{aligned}
	\mathcal{O}_2^{pq} &= \mathcal{O}_{\Lbar QQdH}^{qp}, \\
	\mathcal{O}_1^{pq} & =\mathcal{O}_{\Lbar QQdH}^{pq} -
	\mathcal{O}_{\Lbar QQdH}^{qp} ,
	\label{}
\end{aligned}
\end{equation}
with the lepton flavor index assignment
\begin{equation}
	\mathcal{O}_{\Lbar QQdH}^{pq} = \LC_{ij}(Q^m_p C Q^i_q)(\Lbar_{m}
	d)\Htilde^j.
	\label{}
\end{equation}

All of the dimension-7 operators violate lepton number, and 7 of
the 20 operators violate baryon number as well. In fact, all of the operators
that do not violate baryon number do violate lepton number by two units, i.e.
$L=+2$. The baryon-number violating operators all have $L = -1$, so that $B - L
= 2$; therefore all of the baryon-number violating operators violate $B-L$ as
well.
If the operators violating $B-L$ lead to proton decay, they will be suppressed
by a scale $\Lambda \gtrsim 10^{10}$ GeV, since the proton lifetime is generically
experimentally constrained to be $\gtrsim 10^{32}$ years \cite{Abe:2014mwa, Takhistov:2014pfw,
Abe:2013lua, Regis:2012sn}.  However, the flavor structure of the $B-L$
violating operators could be such that they do not in fact lead to proton decay
within experimentally constrained timescales. This could happen for example for
an operator that did not contain first-generation quarks, and such an operator
might be suppressed by some scale lower than $10^{10}$ GeV.\footnote{I thank an
anonymous reviewer for this insight.}
The operators that do not violate baryon number lead to neutrino mass 
generation, since they have $L= +2$. Therefore these operators are also
suppressed by a high scale, namely $\gtrsim 10^4$ TeV \cite{deGouvea:2007xp}.

As mentioned in the Introduction, some of these dimension-7 operators have
previously been examined in the literature. For example, Ref.~\cite{Babu:2012vb}
lists nine dimension-7 operators with $B = +1$ and ten operators with $L = +2$
in the context of $SO(10)$ grand unified theories and nucleon decay.
The nine $B = +1$ operators are captured by $\mathcal{O}_{\Lbar QddD}^{(1)}$,
$\mathcal{O}_{\Lbar Q ddD}^{(2)}$, $\mathcal{O}_{ddd\ebar D}$,
$\mathcal{O}_{\Lbar QQdH}$, $\mathcal{O}_{\ebar QddH}$, $\mathcal{O}_{\Lbar
dddH}$, and $\mathcal{O}_{\Lbar uddH}$, and the ten $L
= +2$ operators correspond to the remaining five operators in class 6, along
with $\mathcal{O}_{LL\dbar uD}^{(1)}$ (there is not an operator corresponding to
$\mathcal{O}_{LL\dbar uD}^{(2)}$).
Another reference listing
several of the operators is Ref.~\cite{Babu:2001ex}; this deals with $L = +2$
operators that could produce Majorana masses for neutrinos and lists the same
five $L = +2$ operators from class 6 in Table \ref{the_list}. Mention is also made here of
$\mathcal{O}_{LHB}$, $\mathcal{O}_{LHW}$, $\mathcal{O}_{LHDe}$, and the class of
operators comprising $\mathcal{O}_{LHD}^{(1)}$ and $\mathcal{O}_{LHD}^{(2)}$,
but a reduction to the minimal set of operators is not carried out.

\section{Fermions and hypercharge}
\label{fermions_hyper}

It is not possible to construct any dimension-7 operators without using
fermion fields. To see this, recall that in the absence of fermions the
available field content consists only of objects of dimension 1 (the Higgs
doublet $H$ or covariant derivatives $D_\mu$) and objects of
dimension 2 (the field strength tensors $X_{\mu\nu}$). 
Therefore an odd number of dimension-1 objects must be included in order to
obtain a total dimension of 7. However, since the Higgs has hypercharge
$1/2$ and $X_{\mu\nu}$ and $D_\mu$ each have zero
hypercharge, each fermion-free operator must contain an even number of Higgs
fields in order to remain $U(1)_Y$ invariant. Similarly, since the total number
of Lorentz indices in each operator
must be even, each fermion-free operator must contain an even number of
derivatives (since there are no fermions, there are no $\gamma_\mu$'s to provide
another source of Lorentz indices). 
The dimensional constraint requires an odd number of dimension-1 objects and
Lorentz plus hypercharge constraints require an even number of dimension-1
objects, so we can conclude that there are no possible fermion-free operators of
dimension 7, or more generally of any odd dimension.

Furthermore, consider the possibility of dimension-7 operators with multiple
fermion currents. These operators must always have an even number of
fermions since fermion
fields have fractional dimensionality. The maximum possible number of fermion currents is
therefore 2. Since two fermion currents have total dimension 6, we must add a
dimension-1 object in order to get a dimension-7 operator, giving the two classes
$\psi^4 D$ and $\psi^4 H$ discussed in Section \ref{classification}.

Because of the ubiquity of fermions in dimension-7 operators, hypercharge
constraints will play a crucial role in the operator classification in Section
\ref{classification}.
For this reason, the remainder of this section
reviews the basics of fermion currents and hypercharges in the Standard Model and
establishes two facts for easy reference when carrying out the classification.

First, we do not need to use $\gamma_5$ in fermion
currents (such as in the pseudovector matrix $\gamma^\mu \gamma^5$) because
all fermions under discussion are chiral and thus eigenstates of $\gamma_5$. So
the only fermion currents to consider are the scalar, vector, and tensor
currents.
Second, recall that there are two ways to write down a scalar fermion current -
one connecting left-handed fields with right-handed fields and the other
connecting fields of the same chirality with an insertion of charge conjugation,
\begin{equation}
	\psi_{1_{L(R)}} C \psi_{2_{L(R)}}\quad \text{and} \quad
	\overline{\psi}_{1_{L(R)}}
	\psi_{2_{R(L)}} ,
	\label{two}
\end{equation}
where $C$ is the Dirac charge conjugation matrix. The operator $C$ can be explicitly written
as $i \gamma_2 \gamma_0$, and causes a Lorentz spinor to transform as
its conjugate spinor. These two possibilities can also be written in two-component
fermion notation; for a review see Ref.~\cite{Dreiner:2008tw}. In this case it is easiest to
define all fields as
left-handed, so the
Standard Model fermions are $Q, L, u^c, d^c$, and $e^c$, all transforming under
the $(\frac{1}{2},0)$ representation of the Lorentz group (a bar is often used
instead of the superscript $c$ in the field names, but this would further
confuse the notation).
Then the four-component currents $(LCL)$ and $(dCd)$ become
\begin{equation}
	L^\alpha L_\alpha = \LC_{\alpha \beta} L^\alpha L^\beta \quad \text{and}
	\quad d^{c\dagger}_{\dot{\alpha}} d^{c\dagger \dot{\alpha}} =
	\LC^{\dot{\alpha}\dot{\beta}}d^{c\dagger}_{\dot{\alpha}}d^{c\dagger}_{\dot{\beta}} 
	\label{}
\end{equation}
in two-component notation, with $\alpha$ and $\beta$ labelling spinor indices.
Similarly, the four-component current $(\Lbar d)$ is written in two-component
notation as
\begin{equation}
	L^{\dagger}_{\dot{\alpha}} d^{c\dagger \dot{\alpha}} =
	(L_\alpha)^\dagger d^{c\dagger \dot{\alpha}} =
	\LC_{\dot{\alpha}\dot{\beta}} (L^\beta)^\dagger d^{c\dagger\dot{\alpha}}
	.
	\label{}
\end{equation}
Four-component notation is used in the operator list and classification for
continuity with the notation used in Ref.~\cite{Grzadkowski:2010es}, but it is useful
to realize the two-component counterparts, especially when applying Fierz
identities. Appendix \ref{appendix} reproduces the list of dimension-7 operators
from Table \ref{the_list} 
in two-component notation.

Having established some conventions regarding fermions, we now move to the
hypercharge constraints.
Forming all possible scalar fermion currents using the two methods in Eq. \ref{two}
and the fields
from Table \ref{matter_fields} 
shows that there are no scalar fermion currents that have zero
hypercharge. Hereafter this statement will be referred to as {\bf Rule 1}.
The situation for tensor currents mirrors that of scalar currents so far as the 
chirality of the two fields is concerned, so {\bf Rule 1} also applies to tensor
currents.

Similarly, there are two ways to construct a vector fermion current:
\begin{equation}
	\overline{\psi}_{1_{L(R)}}\gamma_\mu \psi_{2_{L(R)}}\quad \text{and}
	\quad \psi_{1_{L(R)}}
	C \gamma_\mu \psi_{2_{R(L)}} .
	\label{}
\end{equation}
Translating to two-component notation gives for example (suppressing the spinor
indices)
\begin{equation}
	L^\dagger \overline{\sigma}^\mu Q \quad \text{and} \quad Q \sigma^\mu
	d^{c\dagger} 
	\label{}
\end{equation}
for the four-component currents $(\Lbar \gamma^\mu Q)$ and $(Q C\gamma_\mu d)$.
Again taking account of all of the possibilities using the fields in Table
\ref{matter_fields}
shows that there are no vector currents with hypercharge $\pm 1/2$. This will be
called {\bf Rule 2}.

Summarizing:
\bigskip
\begin{mydef}
	There are no scalar or tensor fermion currents with zero hypercharge.
\end{mydef}

\bigskip
\begin{mydef}
	There are no vector fermion currents with hypercharge $\pm 1/2$.
\end{mydef}
\bigskip
We will make extensive use of these rules in the following section.

\section{Classification of operators}
\label{classification}

As was done in the previous dimension-6 classifications, the
dimension-7 classification
will use the field equations of motion (EOMs) in order to make
field redefinitions and thus allow some classes of operators to be subsumed into
other classes.
For dimension-7 operators, the EOMs are needed at
$O(1/\Lambda^3)$, and we will neglect $O(1/\Lambda^4)$ effects. Therefore the
EOMs can be calculated using only the original Standard Model Lagrangian
$\mathcal{L}_{SM}$ as given in Eq.~\ref{sm}. It is important to note that this
use of just the classical EOMs is an approximation that may not always
be justified. In particular, it assumes that all of the operators at a given
dimension have the same cutoff scale $\Lambda$. This may not always be the case,
as the particles that are integrated out to give an operator $\mathcal O_a$ may be much
lighter than the particles integrated out to give operator $\mathcal O_b$,
leaving $\mathcal O_a$
with a lower cutoff than $\mathcal O_b$. Hence the cutoff scale $\Lambda$ should
technically have an index $i$ for each independent operator: $\Lambda =\Lambda_i$. As an
example, the dimension-5 Weinberg operator has a cutoff scale $\Lambda =
\Lambda_{\text{dim5}}$ that is
experimentally constrained by the light neutrino masses to be much higher than
many of the experimentally allowed values of $\Lambda = \Lambda_{i,\text{dim6}}$
for dimension-6 operators.

Taking into account the fact that there are no fermion-free operators, the
following 11 distinct classes can be formed from the dimension-7 combinations of
the degrees of freedom $\{X, D, \psi, H\}$:
\begin{equation}
\begin{gathered}
	\psi^2 X^2,\; \psi^2 H^4,\; \psi^2 H^2 D^2,\; \psi^2 H^3 D,\; \psi^2 H
	D^3,\;
	\psi^2 D^4,\; \\ 
	\psi^2 H^2 X,\; \psi^2 D^2 X,\; \psi^2 HDX,\; \psi^4 D,\; \psi^4 H.
	\label{}
\end{gathered}
\end{equation}
These classes are individually examined in the following subsections.
Five classes are completely ruled out by {\bf Rule 1} and {\bf Rule 2}, leaving
the following six classes that will require a closer
look and that do end up containing operators: 1 : $\psi^2 H^4$,
2 : $\psi^2 H^2 D^2$, 3 : $\psi^2 H^3 D$, 4 : $\psi^2 H^2 X$, 5 : $\psi^4 D$,
and 6 : $\psi^4 H$.

Out of these six classes of operators, the EOMs only need to be
used in two classes:
$\psi^2 H^2 D^2$ and $\psi^4 D$. In each of these classes, the use of the
EOMs only reduces some subset of the class of operators to other
classes. This can be contrasted with the situation for the dimension-6 operators,
where the EOMs are instrumental in removing entire classes from
consideration \cite{Grzadkowski:2010es}. Hypercharge constraints are much more
useful for the dimension-7 operators, a fact that can be traced back to the absence
of fermion-free operators of dimension 7.

\subsection{$\psi^2 X^2$}

The total number of Lorentz indices must be even, so the fermion current has to
be a scalar or a tensor. But since the field-strength tensors do not
carry hypercharge, the fermion current must have zero hypercharge, and {\bf
Rule 1} rules this out.

\subsection{$\psi^2 H^4$}

The fermion current must be a scalar and have hypercharge $0$, $\pm 1$, or $\pm
2$. {\bf Rule 1} eliminates the zero hypercharge case. Examining the other
scalar fermion current possibilities using the fields in Table
\ref{matter_fields} shows that only the current with two lepton doublets has
hypercharge $\pm 1$, and only the current with two right-handed electrons has
hypercharge $\pm 2$. 
The lepton doublet current leads to the only operator in
this class
\begin{equation}
	\mathcal{O}_{LH} = \LC_{ij}\LC_{mn} \left( L^i C L^m\right) H^j H^n
	(H^\dagger H).
	\label{}
\end{equation}
The other ways of contracting the $SU(2)_W$ indices are either equivalent or
identically zero.
Forming $SU(2)_W$ triplets instead of singlets does not produce anything new,
because of the group identity
\begin{equation}
	\tau^I_{jk}\tau^I_{mn} = 2 \delta_{jn}\delta_{mk} -
	\delta_{jk}\delta_{mn} .
	\label{}
\end{equation}
For the right-handed electron current, there is no way to contract the $SU(2)_W$
indices in a way that is not identically zero,
\begin{equation}
	(e C e)(H^\dagger_j \Htilde^j)^2 = 
	(e C e) \LC_{ij} \LC_{mn} H^*_i H^*_j H^*_m H^*_n = 0.
	\label{}
\end{equation}

\subsection{$\psi^2 H^2 D^2$}

To form a Lorentz scalar, the fermion current can be either a tensor or scalar,
and it must have
hypercharge $0$ or $\pm 1$. {\bf Rule 1} eliminates the zero hypercharge case, and
the only scalar or tensor current that has hypercharge $\pm 1$ is the one with
two lepton doublets. The remaining analysis in this section closely follows the
calculations done in Ref.~\cite{Grzadkowski:2010es} for the dimension-6 operator class $\psi^2
H D^2$.

Consider first the case with a scalar fermion current. Then there are five
possibilities for where the derivatives act: 1 : both on a single Higgs, 2 :
both on a single fermion, 3 : one on each fermion, 4 : one on each Higgs, and 5
: one on a fermion and the other on a Higgs.
In the work that follows, boxes signify generic classes of operators.
If both derivatives act on a single Higgs, the operator can be reduced using the
Higgs equation of motion
\begin{equation}
	\psi^2 H (D_\mu D^\mu H) = m^2\boxed{\psi^2 H^2} + \boxed{\psi^2 H^4} +
	\boxed{\psi^4 H} \; .
	\label{eom2}
\end{equation}
If both derivatives act on a single fermion, the operator can also be reduced
with EOMs
\begin{equation}
\begin{aligned}
	\psi H^2 (D_\mu D^\mu \psi) &= \psi H^2 (\eta^{\mu\nu}D_\mu D_\nu \psi)
	\\
	& = \psi H^2 \slashed{D}\slashed{D} \psi + \boxed{\psi^2 H^2 X} \\
	& = \boxed{\psi^2 H^3 D} + \boxed{\psi^2 H^2 X} \; .
	\label{eom1}
\end{aligned}
\end{equation}
The second equality in Eq.~\ref{eom1} follows upon using $[D_\mu, D_\nu]\sim
X_{\mu\nu}$ and the identity 
\begin{equation}
	\gamma^\mu \gamma^\nu =
\eta^{\mu\nu} - i \sigma^{\mu\nu},
	\label{gamma_id}
\end{equation}
and the third follows from the fermion EOMs.
If the derivatives act one on each fermion, a combination of integration by
parts and the two previous reductions can be used to change the operator to
possibility number 5:
\begin{equation}
\begin{aligned}
	H^2 (D_\mu \psi)(D^\mu \psi) &= -2 H \psi (D_\mu H)(D^\mu \psi) - \psi H^2
	(D_\mu D^\mu \psi) + \boxed{T}\\
	& = -2H\psi (D_\mu H)(D^\mu \psi) + \boxed{\psi^2 H^3 D} + \boxed{\psi^2
	H^2 X} + \boxed{T}\; ,
	\label{}
\end{aligned}
\end{equation}
where $\boxed{T}$ represents total derivatives, and the second equality follows from
Eq.~\ref{eom1}.
Similarly, if the derivatives act one on each Higgs
\begin{equation}
\begin{aligned}
	\psi^2 (D_\mu H)(D^\mu H) &= -2 H \psi (D_\mu \psi)(D^\mu H) - \psi^2 H
	(D_\mu D^\mu H) + \boxed{T} \\
	& =- 2 H\psi (D_\mu \psi)(D^\mu H) + m^2 \boxed{\psi^2 H^2} +
	\boxed{\psi^2 H^4} + \boxed{\psi^4 H} + \boxed{T} \; ,
	\label{eom3}
\end{aligned}
\end{equation}
where the second equality follows from Eq.~\ref{eom2}. So possibility 4 can also be
reduced to possibility 5.
Now consider the final possibility, number 5, where one derivative acts on a fermion and
the other acts on a Higgs. Integrating by parts in this case just gives back the
same structure, so we need to check if it can be otherwise reduced through gamma
matrix algebra followed by integration by parts:
\begin{equation}
\begin{aligned}
	2\psi H (D^\mu \psi)(D_\mu H) & =2 \psi H (\eta^{\mu\nu}D_\nu \psi)(D_\mu
	H) \\
	& = \psi H ( (\gamma^\mu \slashed{D} + \slashed{D}\gamma^\mu )\psi)(D_\mu
	H) \\
	& = \boxed{\psi^2 H^3 D} - (D_\nu \psi)(\gamma^\nu \gamma^\mu
	\psi)H(D_\mu H) - \psi (\gamma^\nu \gamma^\mu \psi)(D_\nu H)(D_\mu H) \\
	&\quad -
	\psi (\gamma^\nu \gamma^\mu \psi) H (D_\nu D_\mu H) + \boxed{T} \\
	& = \boxed{\psi^2 H^3 D} -\psi^2 (D^\mu H)(D_\mu H) + i \psi
	\sigma^{\mu\nu} \psi (D_\nu H)(D_\mu H) - \psi^2 H (D^\mu D_\mu H) \\
	& \quad + i
	\psi \sigma^{\mu \nu} \psi H(D_\nu D_\mu H) + \boxed{T} \\
	& = \boxed{\psi^2 H^3 D} +\boxed{\psi^2 H^4} +
	\boxed{\psi^4 H} + 
	 m^2 \boxed{\psi^2 H^2} +\boxed{T}  \\
	&\quad +
	2 \psi H (D^\mu \psi)(D_\mu H) + i\psi \sigma^{\mu\nu}\psi (D_\nu
	H)(D_\mu H) + i\psi \sigma^{\mu\nu}\psi (D_\nu D_\mu H) ,
	\label{}
\end{aligned}
\end{equation}
where the penultimate equality follows from the fermion EOMs and Eq.~\ref{gamma_id},
and the final equality follows from Eqs.~\ref{eom2} and \ref{eom3}.
Since we get the same structure back along with some tensor current operators,
this is an independent contribution and
must be included in the operator list.
There are
two independent ways to contract the $SU(2)_W$ indices, since there are four distinct
fundamental representations of $SU(2)_W$ in the tensor product. These give the
following two operators
\begin{equation}
	\mathcal{O}_{LHD}^{(1)} =\LC_{ij}\LC_{mn} L^i C (D^\mu L^j) H^m (D_\mu H^n), 
	\label{op1}
\end{equation}
\begin{equation}
	\mathcal{O}_{LHD}^{(2)} = \LC_{im}\LC_{jn} L^i C (D^\mu L^j) H^m (D_\mu H^n).
	\label{op2}
\end{equation}
The other contraction with $\LC_{in}\LC_{jm}$ is not independent, because of the
Schouten identity
$\LC_{in}\LC_{jm} = \LC_{im}\LC_{jn} - \LC_{ij}\LC_{mn}$.  Note that we can
always choose which fermion the derivative acts on, since the other case is
equivalent up to an integration by parts after using the above results.

Next consider the case with a tensor fermion current. In this case, if both
derivatives act on a single object, since $\sigma_{\mu\nu}$ is antisymmetric we
are led to $[D_\mu, D_\nu]$ and thus to the
class $\psi^2 X H^2$, since $[D_\mu, D_\nu] \sim X_{\mu\nu}$. If one derivative
acts on a fermion and one on a Higgs, by using $\{\gamma^\mu,\gamma^\nu\} =
2\eta^{\mu\nu}$ and the fermion EOMs we have
\begin{equation}
\begin{aligned}
	-(2 i) \psi \sigma_{\mu\nu} (D^\mu \psi)H(D^\nu H) 
	& = \psi (\slashed{D} \gamma_\nu - \gamma_\nu \slashed{D}) \psi H (D^\nu
	H) \\
	& = 2 \psi (D_\nu \psi) H(D^\nu H) - 2 \psi \gamma_\nu (\slashed{D}\psi)
	H (D^\nu H) \\
	& = \boxed{\psi^2 H^3 D} + 2 \psi H (D_\nu \psi)(D^\nu H)  \; ,
	\label{this}
\end{aligned}
\end{equation}
so this possibility can be eliminated in favor of the two operators already
constructed.
If the derivatives act one on each fermion, integrating by parts gives
\begin{equation}
\begin{aligned}
	(D^\mu \psi)\sigma_{\mu\nu} (D^\nu \psi) H^2 & =- \psi \sigma_{\mu\nu}
	H^2 (D^\mu D^\nu \psi) - 2\psi \sigma_{\mu\nu}H (D^\mu \psi)(D^\nu H) +
	\boxed{T} \\
	& = \boxed{\psi^2 X H^2} + \boxed{\psi^2 H^3 D} - 2 i \psi H (D_\nu
	\psi)(D^\nu H) + \boxed{T} \; ,
	\label{}
\end{aligned}
\end{equation}
where the second equality follows from Eq.~\ref{this}. So this possibility can also
be reduced to the operators in Eqs.~\ref{op1} and \ref{op2}.
Similarly, the case with the derivatives acting one on each Higgs reduces to
classes $\psi^2 H^3 D$, $\psi^2 X H^2$, and the previous operators after
integration by parts.  Therefore there
are no tensor current operators in this class.

\subsection{$\psi^2 H^3 D $}

The fermion current must be a Lorentz vector with hypercharge $\pm 1/2$ or $\pm
3/2$. {\bf Rule 2} eliminates the hypercharge $\pm 1/2$ case, and the only
remaining vector current that will work is the one connecting the lepton doublet
with the right-handed electron field.

If the derivative acts on either of the fermion fields, two identical Higgs
doublets must be contracted in order to satisfy $SU(2)_W$ invariance, giving a
result that is identically zero.
If the derivative does not act on either fermion field, we only have to consider
the derivative acting on a single Higgs doublet, since all of the remaining
cases reduce to this. Then there is only one way to contract the $SU(2)_W$
indices that is not identically zero, giving the operator
\begin{equation}
	\mathcal{O}_{LHDe} =
	\LC_{ij}\LC_{mn}\left(L^i C\gamma_\mu e\right) H^j H^m D^\mu H^n .
	\label{}
\end{equation}

\subsection{$\psi^2 H D^3$}

The fermion current must be a Lorentz vector with hypercharge $\pm 1/2$. But
this possibility is removed by {\bf Rule 2}.

\subsection{$\psi^2 D^4$}

The fermion current must be a scalar or tensor with hypercharge zero, and
therefore {\bf Rule 1} eliminates this class.

\subsection{$\psi^2 H^2 X$}

The fermion current must be a tensor with hypercharge 0 or $\pm 1$, since the
field-strength tensors are traceless. The only
current that works is the one with two lepton doublets.  
For $B_{\mu\nu}$ 
there is only
one independent way to contract the $SU(2)_W$ indices, giving the following
operator
\begin{equation}
	\mathcal{O}_{LHB} = \LC_{ij}\LC_{mn}(L^i C
	\sigma_{\mu\nu} L^m) H^j H^n B^{\mu\nu} ,
	\label{}
\end{equation}
For $W_{\mu\nu}^I$, we need to include the triplet $\tau^I$, and there are
generally six
$SU(2)$ singlets in the product of four fundamentals and two triplets. However,
some cases vanish because of the two identical $H$ fields, leaving
four independent singlets. Then allowing family index transpositions between the
two leptons (as in the examples in Section \ref{list}) and accounting for the
fact that the labels on the Higgs fields are
interchangeable leaves only one independent operator:
\begin{equation}
	\mathcal{O}_{LHW} = \LC_{ij}\left(\tau^I \LC\right)_{mn} (L^i C
	\sigma_{\mu\nu} L^m) H^j H^n W^{I\mu \nu} .
	\label{}
\end{equation}

The gluon field-strength tensor
$G^A_{\mu\nu}$ cannot be used since there are no other objects available with a nontrivial
$SU(3)_C$ transformation.
The possibility of using the dual tensors $\widetilde{B}^{\mu\nu}$ or
$\widetilde{W}^{I\mu\nu}$ does not give any new operators, because of the
identities
\begin{equation}
	\LC_{\alpha\beta\mu\nu}\sigma^{\mu\nu} = 2 i \sigma_{\alpha\beta}
	\gamma_5
	\label{id1}
\end{equation}
and
\begin{equation}
	\gamma_5 \psi_{L,R} = \mp \psi_{L,R}.
	\label{id2}
\end{equation}

\subsection{$\psi^2 D^2 X$}

The fermion current must be a scalar or tensor with zero hypercharge, so this
class is eliminated by {\bf Rule 1}.

\subsection{$\psi^2 H D X$ }

The fermion current must be a vector with hypercharge $\pm 1/2$, so {\bf Rule 2}
removes this class.

\subsection{$\psi^4 D$}

The fermions must have zero total hypercharge, and one of the two currents must
be a vector current.
If the derivative acts on either of the two fermions in the vector current, the
operator can be
reduced by using the fermion EOMs to the class $\psi^4 H$.
If the derivative acts on the scalar or tensor current, the operator cannot be
reduced. There are only five current combinations that have a single vector current
and zero total hypercharge: $(\dbar L)(LC\gamma_\mu u)$,
$(\Lbar d)(QC\gamma_\mu d)$, $(\Lbar \gamma_\mu Q)(dCd)$, $(\dbar \gamma_\mu
u)(LCL)$, and
$(dCd)(\ebar \gamma_\mu d)$. Using integration by parts
along with the fermion EOMs allows the derivative to be switched back and forth
between the two fermions in the scalar or tensor current
\begin{equation}
\begin{aligned}
(\psi_1 \gamma_\mu \psi_2)( (D^\mu \psi_3) \psi_4) & = (\psi_1 \gamma_\mu
 \psi_2)(\psi_3 D^\mu \psi_4) + (D^\mu \psi_1 \gamma_\mu \psi_2)(\psi_3 \psi_4)
 \\
 & \qquad + (\psi_1 \gamma_\mu D^\mu \psi_2)(\psi_3 \psi_4) + \boxed{T}  \\
 & = (\psi_1 \gamma_\mu \psi_2)(\psi_3 D^\mu \psi_4) + \boxed{\psi^4 H} +
 \boxed{T} \; , 
\end{aligned}
\end{equation}
so we can arbitrarily pick one of the two fermions for the derivative to act on.
Doing this for each of the five current combinations gives the
following five operators:

\begin{equation}
	\mathcal{O}_{LL\dbar uD}^{(1)} =\LC_{ij} (\dbar \gamma_\mu u)(L^i CD^\mu
	L^j),
	\label{}
\end{equation}
\begin{equation}
	\mathcal{O}_{LL\dbar u D}^{(2)}  =\LC_{ij} (\dbar \gamma_\mu u)(L^i C
	\sigma^{\mu\nu}
	D_\nu L^j),
	\label{}
\end{equation}
\begin{equation}
	\mathcal{O}_{\Lbar Q dd D}^{(1)} = (Q C \gamma_\mu d)(\Lbar D^\mu d),
	\label{}
\end{equation}
\begin{equation}
	\mathcal{O}_{\Lbar Q dd D}^{(2)} = (\Lbar \gamma_\mu Q)(d C D^\mu d),
	\label{}
\end{equation}
\begin{equation}
	\mathcal{O}_{ddd\ebar D} = (\ebar \gamma_\mu d)(d C D^\mu d).
	\label{}
\end{equation}
The tensor current was chosen for $\mathcal{O}_{LL\dbar uD}^{(2)}$ to avoid
crossing color indices between currents. The scalar current not included can
then be formed by using Fierz identities and flavor index transpositions of
$\mathcal{O}_{LL\dbar uD}^{(1)}$ and $\mathcal{O}_{LL\dbar uD}^{(2)}$. See
Ref.~\cite{Dreiner:2008tw} or Ref.~\cite{Bailin:1994qt} for the relevant Fierz identities.

\subsection{$\psi^4 H$}

The fermions must have total hypercharge $\pm 1/2$. 
The operator must be constructed from two scalar
currents, two vector currents, or two tensor currents in order to preserve Lorentz
invariance. This places constraints on
the hypercharge combinations that actually work.

For a given field content, we can calculate the number of $SL(2,\mathbb{C})$
singlets in the tensor product, thus allowing a direct statement of the number
of independent operators with that field content without an explicit calculation
of all the possible Fierz transformations. For example, the field content $\{L,
L, \Qbar, u\}$, can be written as the product of two vector currents or as the
product of two scalar currents, but there is only one $SL(2,\mathbb{C})$ singlet
in the tensor product
$(\frac{1}{2}, 0)\otimes (\frac{1}{2}, 0)\otimes (0,\frac{1}{2})\otimes
(0,\frac{1}{2})$.  So
we can choose the product of scalar currents as the representative operator (this
choice does not cross color indices between currents, so it is aesthetically more
pleasing). 
Now we need to consider the $SU(2)_W$ contraction. There are three ways to
contract the $SU(2)_W$ indices, two of which are independent. But since there
are two identical fields in the operator, two of the three $SU(2)_W$
contractions are equivalent under a transposition of flavor indices. Therefore there
is only one independent operator for this set of fermion fields:
\begin{equation}
	\mathcal{O}_{LL\Qbar uH} = \LC_{ij} (\Qbar_m u)(L^m C L^i)H^j .
	\label{}
\end{equation}

The field contents that work for this class can be found by a simple search, and
they are listed in Table \ref{fermions}. Carrying out the procedure described in
the previous paragraph for these eight sets of fermion fields leads to the
operators listed for this class in Table \ref{the_list}. Whenever possible the
currents are chosen so that color indices are not crossed between currents.
Note that for the field content $\{L, L, Q,
\dbar, H\}$, there are two $SL(2,\mathbb{C})$ singlets and two ways to contract
the $SU(2)_W$ indices for each of these singlets, leading to two independent
operators after considering possible flavor index permutations. In this case,
the color indices must be crossed between currents 
so that all of the  possible
$SU(2)_W$ singlets and Lorentz contractions can be formed from the two operators
given.

\begin{table}
	\begin{center}
		{
			\begin{tabular}{cc}
			\toprule
			\multicolumn{2}{c}{Fields} \\
			\midrule
			$\{L,L,\Qbar,u\}$	&	$\{\Lbar,u,d,d\}$	\\
			$\{\Lbar,Q,Q,d\}$	&	$\{\Lbar,d,d,d\}$	\\
			$\{\ebar,Q,d,d\}$	&	$\{L,L,Q,\dbar\}$	\\
			$\{L,e,u,\dbar\}$	&	$\{L,L,L,\ebar\}$	\\
		        \bottomrule
		\end{tabular}}
		\caption{The eight sets of four fermion fields that can be joined
			into two scalar or two vector currents, have a total
			hypercharge $\pm 1/2$, and have an odd number of
			$SU(2)_W$ doublets. The entries in the first column
			allow a single $SL(2,\mathbb{C})$ singlet, and the
			entries in the second column give two $SL(2,\mathbb{C})$
			singlets.}
		\label{fermions}
	\end{center}
\end{table}

\section{Conclusion}
\label{conclusion}

The Standard Model works extremely well at explaining particle physics as we know it,
and no clear BSM signals have come into view at the LHC 7 and 8 TeV runs.  
It is therefore imperative to study in detail all possible deviations from the
Standard Model in order to better understand the specific channels where new
physics might materialize.
To further this program, we have presented a complete classification of the
dimension-7 operators in the Standard Model Effective Field Theory. 
There are 20 dimension-7 operators, all lepton number violating, with 7 of them also
violating baryon number. All of the operators include fermions, so the use of
hypercharge constraints plays a central role in the operator classification.
This
catalogue allows a closer examination of the SMEFT structure and
properties, and provides a guide for detailed phenomenological studies utilizing an
effective field theory approach to physics beyond the Standard Model. Even
though most of the operators are generically suppressed by a very high scale, it
would be interesting to try to find loopholes in this supression, perhaps by
utilizing the flavor structure of the $B-L$ operators as mentioned in Section
\ref{list}.

Some simple modifications allow the construction of more independent operators;
extending the reach of the SMEFT beyond the Standard Model.
For example,
adding another distinct Higgs doublet, as is done in supersymmetry, allows the
construction of a new operator at the dimension-5
level,\footnote{This operator is $\LC_{ij}\LC_{mn}(L_p^i C L_q^j)H_1^mH_2^n $.} 
and would give several
new operators at dimensions 6 and 7. Modifying the Standard Model by including
right-handed neutrinos also gives many new operators, some of which
were examined at the dimension-6 level in Ref.~\cite{Alonso:2014zka}.
The SMEFT could also be extended while remaining strictly within the confines of
the Standard Model by perfoming a classification of the
dimension-8 operators. This would certainly be possible but would be tedious
considering the large number of operators available at dimension 8 as opposed to 
dimension 7.

It would also be interesting to calculate the one-loop anomalous dimension
matrix for the
dimension-7 operators and check what, if any, of the holomorphy properties
defined in Ref.~\cite{Alonso:2014rga} are
present, but this is beyond the scope of this work. At this point, we simply
note that, according to the definition of holomorphy given in Ref.~\cite{Alonso:2014rga}, 
there are 10 holomorphic and anti-holomorphic operators, and
10 nonholomorphic operators at
dimension 7. Since none of the dimension-7 operators are self-conjugate, the
hermitian conjugates of the holomorphic (anti-holomorphic) operators will be
anti-holomorphic (holomorphic), and the hermitian conjugates of the
nonholomorphic operators will also be non-holomorphic. Perhaps some new
structure may emerge when the formal properties of the Standard Model effective
field theory are examined in more detail.

\acknowledgments

I thank Joseph Bramante and Adam Martin for illuminating discussions.

\appendix
\section{The operators in 2-component notation}
\label{appendix}
\begin{table}[H]
\begin{center}
\begin{minipage}[t]{0.45\textwidth}
\renewcommand{\arraystretch}{1.5}
\begin{tabular}[t]{c|c}
\multicolumn{2}{c}{1 : $\psi^2 H^4$ + h.c.} \\
\hline
$ \mathcal{O}_{LH}$ 	& $\LC_{ij}\LC_{mn}(L^i L^m)H^j H^n (H^\dagger H)$ \\
\end{tabular}
\end{minipage}
\begin{minipage}[t]{0.45\textwidth}
\renewcommand{\arraystretch}{1.5}
\begin{tabular}[t]{c|c}
\multicolumn{2}{c}{2 : $\psi^2 H^2 D^2$ + h.c.} \\
\hline
$ \mathcal{O}_{LHD}^{(1)} $ & $\LC_{ij}\LC_{mn} (L^i D^\mu L^j)H^m D_\mu H^n$\\
$ \mathcal{O}_{LHD}^{(2)} $ & $\LC_{im}\LC_{jn} (L^i D^\mu L^j)H^m D_\mu H^n$\\
\end{tabular}
\end{minipage}
\vspace{0.25cm}
\begin{minipage}[t]{0.45\textwidth}
\renewcommand{\arraystretch}{1.5}
\begin{tabular}[t]{c|c}
\multicolumn{2}{c}{3 : $\psi^2 H^3 D$ + h.c.} \\
\hline
$\mathcal{O}_{LHDe}$ 	& $\LC_{ij}\LC_{mn}\left(L^i \sigma_\mu \ebar^\dagger \right) H^j H^m
D^\mu H^n$  \\
\end{tabular}
\end{minipage}
\begin{minipage}[t]{0.45\textwidth}
\renewcommand{\arraystretch}{1.5}
\begin{tabular}[t]{c|c}
\multicolumn{2}{c}{4 : $\psi^2 H^2 X$ + h.c.} \\
\hline
$\mathcal{O}_{LHB}$ 	& $\LC_{ij}\LC_{mn}\left(L^i \sigma_{\mu\nu}L^m \right)
H^j H^n B^{\mu\nu}$ \\
$\mathcal{O}_{LHW}$ 	& $\LC_{ij}(\tau^I \LC)_{mn} \left(L^i  \sigma_{\mu\nu} L^m\right)
H^j H^n W^{I\mu\nu}$ \\
\end{tabular}
\end{minipage}
\vspace{0.25cm}
\begin{minipage}[t]{0.45\textwidth}
\renewcommand{\arraystretch}{1.5}
\begin{tabular}[t]{c|c}
\multicolumn{2}{c}{5 : $\psi^4 D$ + h.c.} \\
\hline
$\mathcal{O}_{LL\dbar uD}^{(1)}$ & $\LC_{ij}(\dbar \sigma_\mu \ubar^\dagger )(L^i  D^\mu
L^j)$ \\
$\mathcal{O}_{LL\dbar uD}^{(2)}$ & $\LC_{ij}(\dbar \sigma_\mu \ubar^\dagger
)(L^i  \sigma^{\mu\nu} D_\nu L^j)$ \\
$\mathcal{O}_{\Lbar QddD}^{(1)}$ & $(Q\sigma_\mu \dbar^\dagger )(L^\dagger D^\mu
\dbar^\dagger ) $ \\
$\mathcal{O}_{\Lbar QddD}^{(2)}$ & $(Q \sigma_\mu L^\dagger )(\dbar^\dagger D^\mu
\dbar^\dagger) $ \\
$\mathcal{O}_{ddd\ebar D} $	& $ (\ebar \sigma_\mu \dbar^\dagger
)(\dbar^\dagger D^\mu \dbar^\dagger )$ \\
\end{tabular}
\end{minipage}
\begin{minipage}[t]{0.45\textwidth}
\renewcommand{\arraystretch}{1.5}
\begin{tabular}[t]{c|c}
\multicolumn{2}{c}{6 : $\psi^4 H$ + h.c.} \\
\hline
$\mathcal{O}_{LLL\ebar H} 	$	& $\LC_{ij}\LC_{mn}(\ebar L^i)(L^j
L^m)H^n$  \\
$\mathcal{O}_{LLQ\dbar H}^{(1)} $ & $\LC_{ij}\LC_{mn}(\dbar L^i)(Q^j L^m)H^n$ \\
$\mathcal{O}_{LLQ\dbar H}^{(2)} $ & $\LC_{im}\LC_{jn}(\dbar L^i)(Q^j L^m)H^n$
\\
$\mathcal{O}_{LL\Qbar uH} $	& $\LC_{ij}(Q^{\dagger}_m \ubar^\dagger)(L^m L^i)H^j$	\\
$\mathcal{O}_{\Lbar QQdH}  	$	& $\LC_{ij}(L^{\dagger}_m \dbar^\dagger )(Q^m 
Q^i)\Htilde^j$ \\
$\mathcal{O}_{\Lbar dddH} $ 	& $(\dbar^\dagger \dbar^\dagger)(L^\dagger
\dbar^\dagger)H$ \\
$\mathcal{O}_{\Lbar uddH} $ 	& $(L^\dagger \dbar^\dagger)(\ubar^\dagger
\dbar^\dagger)\Htilde$ \\
$\mathcal{O}_{Leu\dbar H} $ 	& $\LC_{ij}(L^i \sigma_\mu \ebar^\dagger)(\dbar
\sigma^\mu \ubar^\dagger )H^j$ \\
$\mathcal{O}_{\ebar QddH} $ 	& $\LC_{ij}(\ebar Q^i)(\dbar^\dagger
\dbar^\dagger)\Htilde^j$ \\
\end{tabular}
\end{minipage}
\end{center}
\caption{The dimension-seven operators in 2-component fermion notation. 
	All of the fermions are defined to be fundamentally left-handed, so the
	fermion fields are $Q, L, \dbar, \ubar$, and $\ebar$, all transforming under
	the $(\frac{1}{2}, 0)$ representation of the Lorentz group. The bar here
	is part of the field name and in particular does \emph{not} mean the Dirac bar
	used in 4-component notation. 
	Spinor indices are contracted within parentheses.
	Color
	and flavor indices are left
	implicit, and 
	$SU(2)_W$ indices are left implicit when the contractions are
	obvious. The symbol $C$ represents the Dirac charge conjugation matrix,
	as explained in Section \ref{fermions_hyper}.}
\end{table}

\bibliography{Notes}
\bibliographystyle{jhep}

\end{document}